\documentclass[preprint]{aastex}

\usepackage{natbib}
\shorttitle{J093010} 
\shortauthors{J.-R. KOO ET AL.} 

\begin{document}
\title{1SWASP J093010.78+533859.5: A Possible Hierarchical Quintuple System} 
\author{Jae-Rim Koo$^{1,2}$, Jae Woo Lee$^1$, Byeong-Cheol Lee$^1$, Seung-Lee Kim$^1$, Chung-Uk Lee$^1$, \\
Kyeongsoo Hong$^1$, Dong-Joo Lee$^1$, and Soo-Chang Rey$^2$}
\affil{$^1$Korea Astronomy and Space Science Institute, Daejeon 305-348, Korea;\email{koojr@kasi.re.kr}}
\affil{$^2$Department of Astronomy and Space Science, Chungnam National University,\\ Daejeon 305-764, Korea}

\begin{abstract}
Among quadruples or higher multiplicity stars, only a few doubly eclipsing binary systems have been discovered.
They are important targets to understand the formation and evolution of multiple stellar systems
because we can obtain accurate stellar parameters from photometric and spectroscopic studies.
We present the observational results of this kind of rare object 1SWASP J093010.78+533859.5,
for which the doubly eclipsing feature had been detected previously from the SuperWASP photometric archive.
Individual PSF photometry for two objects with a separation of about 1.9 arcsec was performed for the first time in this study. 
Our time-series photometric data confirms the finding of \citet{lohr2013} that the bright object A is an Algol-type detached eclipsing binary and
the fainter B is a W UMa-type contact eclipsing.
Using the high-resolution optical spectra, we obtained well-defined radial velocity variations
of system A. Furthermore, stationary spectral lines were detected that must have originated from 
a further, previously unrecognized stellar component.
It was confirmed by the third object contribution from the light curve analysis.
No spectral feature of the system B was detected, probably due to motion blur by long exposure time.
We obtained the binary parameters and the absolute dimensions of the systems A and B from light curve synthesis with and without radial velocities, respectively. 
The primary and secondary components of system A have a spectral type of K1 and K5 main sequences,
respectively. Two components of system B have nearly the same type of K3 main sequence.
Light variations for both binaries are satisfactorily modeled by using two-spot models with one starspot on each component.
We estimated the distances to systems A and B individually. 
Two systems may have similar distances of about 70 pc and seem to be gravitationally bound
with a separation of about 130 AU.
In conclusion, we suggest that 1SWASP J093010.78+533859.5 is a quintuple stellar system
with a hierarchical structure of a triple system A(ab)c and a binary system B(ab).

\end{abstract}

\section{Introduction}
A stellar system consists of a small number of stars that are gravitationally bound and revolve each other around their center of mass.
Since the stars in such a system are born nearly at the same time within a given space and evolve with interaction each other,
they are important objects for understanding the stellar formation and dynamical evolution.
The updated multiple star catalogue \citep[MSC;][]{tokovinin1997} contains totally 1,359 stellar systems of multiplicity 3 to 7 in 2010  (at http://www.ctio.noao.edu/\textasciitilde{}atokovin/).
Among the most frequent triple stars, triply eclipsing hierarchical triple systems are very rare and fascinating to show 
peculiar light variations caused by the distant third component passing across the close eclipsing binary. 
To our knowledge, these tertiary eclipsing signals have been detected only in three systems, HD 181068 \citep{derekas2011},
KOI-126 \citep{carter2011}, and KIC 2856960 \citep{armstrong2012,jwlee2013}, all from the $Kepler$ space telescope.

Another interesting object is a doubly eclipsing multiple system, which consists of at least two eclipsing binaries.
We can investigate their physical characteristics in detail because the stellar mass and radius for each component are 
accurately and directly derived from both spectroscopic and photometric observations. 
They are also rare and only a handful of systems have been discovered (\citealp[ADS 9537AB,][]{batten1965}; \citealp[V994 Her,][]{leecu2008}; 
\citealp[OGLE LMC-ECL-16549,][]{graczyk2011}; \citealp[KIC 4247791,][]{lehmann2012}; \citealp[CzeV343,][]{cagas2012}).
On the other hand, \citet{ofir2008} detected two eclipsing light variations with periods of about 3.58 and 5.37 days
in the object OGLE J051343.14$-$691837.1 (= BI 108), which is located at the open cluster NGC 1881 in the Large Magellanic Cloud.
He suggested that the object is a compact hierarchical system of two eclipsing binary pairs in a 3:2 resonance. 
Later, \citet{kolaczkowski2013} performed spectroscopic observations and confirmed radial velocity variations by the longer period binary.
However, unexpectedly no spectral feature by the shorter period binary was detected.
Although its origin of variability may remain unclear, they suggested that the most plausible hypothesis is a doubly eclipsing quadruple system
with the binary orbital period ratio of precise 3:2 by chance or by unknown dynamical reasons.

Recently, \citet{lohr2013} reported the discovery of a new doubly eclipsing multiple system 
1SWASP J093010.78+533859.5 (hereafter J093010) from the SuperWASP \citep[Wide Angle Search for Planets,][]{pollacco2006} archive.
The folded light curves with two periods ($\sim$0.2277 and $\sim$1.3055 days) show typical shapes of the contact and detached eclipsing binaries, respectively.
The J093010 has been known to be a visual binary, TYC 3807-759-1 (hereafter J093010A) and TYC 3807-759-2 (hereafter J093010B), separated by 1.9 arcsec.
Although the SuperWASP data could not resolve these two objects, \citet{lohr2013} suggested that the light variations originated from a
doubly eclipsing quadruple system. They calculated the separation between the two binaries to be 66.1 AU, by applying the distance of 35 pc
estimated from the  photometric parallax by \citet{agueros2009}, and derived their meta-orbital period to be $\sim$400 years.
It should be noted that the proper motion from the Tycho was measured at the photocenter of J093010 \citep[][]{fabricius2002}, 
not for two individual objects. It may still be uncertain, therefore, whether the two objects are gravitationally bound. 

In this study, we investigated the hierarchical structure of the multiple system J093010 and the physical properties of each component in detail 
from the follow-up spectroscopic and photometric observations.
In Section 2, we describe the follow-up observations.
The data analysis and modeling are represented in Sections 3 and 4, respectively.
Finally, the summary and discussion are given in Section 5.

\section{Observations}

\begin{figure}
\includegraphics[]{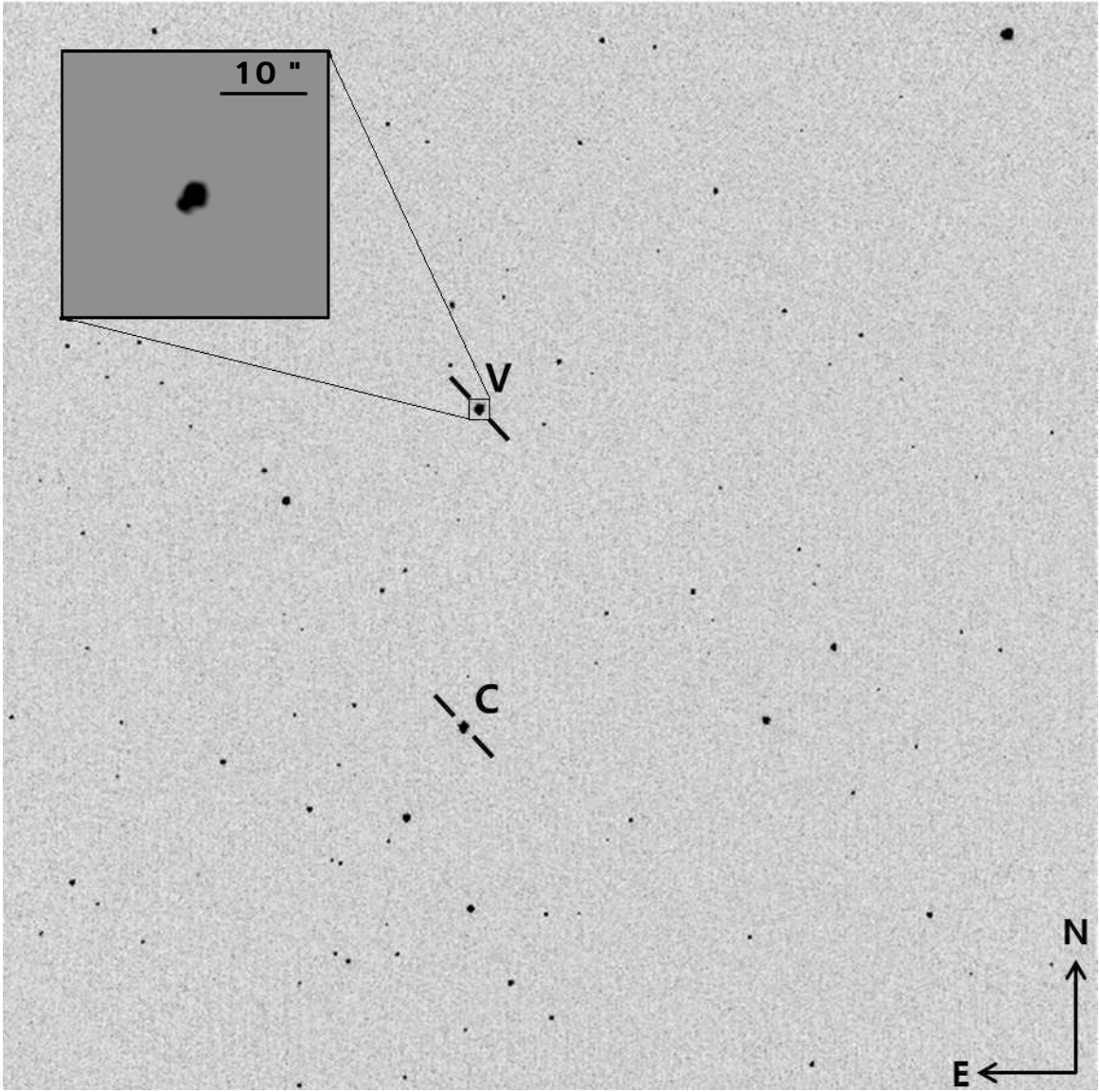}
\caption{$V$ band observed image with 28.1 $\times$ 28.1 arcmin$^2$ field of view.
J093010 and a comparison star (TYC 3807-54-1) are marked as `V' and `C', respectively.
The zoomed box image with about 30 $\times$ 30 arcsec$^2$ taken during the spectroscopic observations shows two objects, being separated by about 1.9 arcsec.
The brighter object is TYC 3807-759-1 (J093010A) and the fainter is TYC 3807-759-2 (J093010B).
\label{Fig1}}
\end{figure}

High-resolution spectroscopic observations were performed to measure the radial velocity (RV) of each component.
Spectra of J093010 were obtained from 2012 November to 2013 March
by using the fiber-fed spectrograph BOES \citep[Bohyunsan Optical Echelle Spectrograph;][]{kim2007} 
attached to the 1.8 m reflector at BOAO (Bohyunsan Optical Astronomy Observatory) in Korea.
The BOES has a long-term stability of about 6.8 m s$^{-1}$ by observing the RV standard star $\tau$ Ceti \citep{bclee2012}.
Since the object is not bright enough to get high signal-to-noise ratio (SNR) spectra, we selected the 2$\times$2 binning mode and the largest fiber with a diameter of 300 $\mu m$, corresponding to a field of view (FOV) of 4.3 arcsec and a resolving power of R = 30,000. 
The observed spectra could be contaminated by sky light or neighbouring stars due to a large size fiber.
But there is no bright object within 1.0 arcmin (see Figure \ref{Fig1}) and the sky was not bright enough to contaminate the spectra during observations.
Exposure time was mostly set to be 20 minutes considering the period for the contact binary and the luminosity of fainter object. 
However, it is not short enough to avoid motion blur of the spectral lines for the contact binary. 
We additionally obtained one spectrum for each object by using the smallest fiber of 80 $\mu m$ diameter (R = 90,000) with 1.1 arcsec FOV on 
3$\times$3 binning mode under good weather conditions.
The resulting SNR of the spectra at the region of $\sim$5500$\rm{\AA}$ was around 30.
Th-Ar Arc and Tungsten-Halogen lamps were taken for wavelength calibration and flat correction.

Photometric observations were carried out using the 4k CCD camera attached to a 1.0 m telescope at LOAO (Mount Lemmon Optical Astronomy Observatory) in Arizona, USA, 
in 2012 December and 2013 January/February. Detailed information of the telescope and detector can be found in \citet{lee2012}. The logs for spectroscopic and photometric observations are listed in Table \ref{tab0}.

\section{Data Analysis}

The basic reductions such as bias subtraction and flat field correction were performed for spectroscopic and photometric data using IRAF packages.
Figure \ref{Fig1} displays a sample of $V$ band images and a zoomed image of J093010 to show two objects with a separation of 1.9 arcsec.
We applied PSF (Point Spread Function) fitting photometry using IRAF/DAOPHOT to obtain instrumental magnitudes for each object.
Then, differential photometry was performed with TYC 3807-54-1 as a comparison star, which was selected by considering the color, brightness and constancy in apparent light. 
Table \ref{tab1} lists the results of differential photometry.
We classified that the brighter object J093010A is an Algol-type and the fainter J093010B is a W UMa-type eclipsing binary.
The folded light curves of each object are shown in Figure \ref{Fig2}. In this paper, we excluded data points that showed an unreasonably large deviation compared to neighbouring ones.

The times for the eclipsing minimum and their errors were determined via the method of \citet{kwee1956}, and listed in Tables \ref{tab2} and \ref{tab3}.
By introducing all minimum times in the tables into a linear least-squares fit, we determined the light ephemeris of our observations for 
J093010A and J093010B to be
\begin{equation}\label{eq1}
{\rm Min. ~I} = {\rm HJD ~} 2456346.7849(15)+ 1.305502(41)E
\end{equation}
\begin{equation}\label{eq2}
{\rm Min. ~I} = {\rm HJD ~} 2456267.92908(22)+ 0.2277135(16)E,
\end{equation}
respectively, where $E$ is the number of orbital cycles that elapsed from the reference epoch and the parenthesized numbers are 
the standard errors for the last digits of each term of the ephemeris.
We independently derived each light element from light-curve synthesis, which used all data points and included spot parameters, in the following section. 
The periods of each binary are almost same with those presented by \citet{lohr2013} in spite of our shorter time span.

We detected three components from the observed spectra of J093010, which were easily identified at Figure \ref{Fig3}. 
Broad lines from two components shifted blueward or redward periodically with an orbital phase of J093010A while the other relatively sharp lines stayed to be constant.
At the H$\alpha$ 6563 \AA\, region, the emissions are shown from the secondary component of J093010A, but clear signals of either the absorption or 
emission are absent for the primary component.
These features might be the result of the quantitatively balanced emission with the absorption of the primary component.
In the spectra, any line for the fainter object J093010B was not detected.
We tried to identify the components from each spectrum that was obtained with a smaller fiber (80 $\mu m$).
Three components were detected in the spectrum of J093010A. However, it was difficult to identify any component in 
the spectrum of J093010B, probably due to its rotational broadening and motion blur by relatively long exposure time.
This indicates that all three components detected from the time-series spectra are members of the brighter object J093010A.

We applied the line depth ratio (LDR) to estimate the temperature for the stationary component. 
\citet{catalano2002} presented many correlations between the LDRs of line pairs and effective temperatures.
Among them, we chose the line pairs of $\lambda$6199 V ${\rm I}$ -- $\lambda$6200 Fe ${\rm I}$, 
$\lambda$6252 V ${\rm I}$ -- $\lambda$6253 Fe ${\rm I}$, and $\lambda$6266 V ${\rm I}$ -- $\lambda$6265 Fe ${\rm I}$ with a relatively tight correlation.
Selected spectra from the $\lambda$6200 region were combined to avoid the blending of J093010A.
The averaged effective temperature of T$_{\rm{eff}} \approx$  4,760 $\pm$ 100 K was derived.

Synthetic spectra were calculated using the stellar atmospheric model of \citet{kurucz1993} and the GNU-Linux version 
of the ATLAS9/SYNTHE code given by \citet{sbordone2004} and \citet{sbordone2005}.
The synthetic spectrum, adopting the solar metal abundance, T$_{\rm{eff}}$ = 4,750 K, log $g$ = 4.5, and $v\sin i$ = 4 km s$^{-1}$, 
was comparable to the observed one of the stationary component.
The LDR for the other two components was not suitable because the absorption lines were broad and blended with each other.
The observed spectra of the primary component appeared to be similar to the synthetic spectrum of T$_{\rm{eff}}$ $\approx$ 5,000 K.
The temperature is consistent with the color index discussed in the following section.

Although the cross-correlation function (CCF) technique is a standard method to determine the RV, we could not examine the CCF 
because the primary and secondary spectra were broadened by their rotations and blended with each other in almost all phases. 
Instead, line-profile fitting using three Gaussian functions was applied to the isolated absorption lines ($\lambda$6431, $\lambda$6439 and $\lambda$6678 Fe ${\rm I}$, $\lambda$6644 Ni ${\rm I}$ and $\lambda$6718 Ca ${\rm I}$). 
The fitting was performed with MPFIT \citep{markwardt2009}.
Figure \ref{Fig4} displays selected samples of fitting for $\lambda$6678 Fe ${\rm I}$ lines.
Since the spectra near eclipses are heavily blended with each other, our measured RVs near the phase of $\phi \sim$ 0.5 contain large uncertainties (Figure \ref{Fig5}).
The averaged RVs and their standard deviations of the each component are listed in Table \ref{tab4} and displayed in Figure \ref{Fig5} with the fitted model velocities. The system velocity ($\gamma_{0}$) and semi-amplitudes ($K_{\rm 1}$ and $K_{\rm 2}$) of the velocity curve were measured to be $-$10.1, 99.4, and 118.2 km s$^{-1}$, respectively.
The RVs of the third component are almost constant ($-$7.1 $\pm$ 0.6 km s$^{-1}$) within their errors. 

\section{Data Modeling}

Light-curve syntheses of J093010A and J093010B were performed using the 2003 version of the Wilson-Devinney synthesis code 
\citep[][hereafter W-D code]{wilson1971}. From the Tycho-2 Catalogue \citep{hog2000}, the color indices are given 
as ($B-V$)$_{\rm A}$= +0.85 $\pm$ 0.05 for J093010A and ($B-V$)$_{\rm B}$ = +0.83 $\pm$ 0.13 for J093010B. The Tycho data 
may not be accurate due to the blending of two systems and there is no information of orbital phases for the measurements. 
Therefore, we tried to get more accurate magnitudes and color indices for each object by using our photometric data.
In order to normalize instrumental magnitudes, we applied an ensemble normalization technique \citep{gilliland1988} following 
the same procedure as that used by \citet{kim2001}. Secondary standard stars were selected from the UCAC4 catalogue \citep{zacharias2013}.
Table \ref{tab5} lists the standard magnitudes and color indices at the maxima and minima of two eclipsing binaries, 
and their errors are standard deviations for the measurements around the selected phases. The maxima of J093010A are located 
near the phase 0.55. The de-reddened colors are derived after adopting $E$($B-V$) = $+$0.01 following \citet{schlafly2011}.

We assumed the gravity-darkening exponents and bolometric albedoes to be $g$ = 0.32 \citep{lucy1967} and $A$ = 0.5 \citep{rucinski1969}, 
respectively, which is appropriate for stars with convective envelopes. The square root bolometric ($X$, $Y$) and 
monochromatic ($x$, $y$) limb-darkening coefficients were interpolated from the values of \citet{van1993} in concert with 
the model atmosphere option. Furthermore, spot models were added to fit light variations and the third light source ($l_3$) 
was considered throughout the analyses. The light-curve syntheses for two binaries were repeated until the correction of each parameter become smaller 
than its standard deviation by using the differential correction (DC) program of the W-D code. The values with parenthesized errors 
in Tables \ref{tab6} and \ref{tab7} signify adjusted parameters. The subscripts 1 and 2 refer to the primary and secondary stars eclipsed 
at Min I (at phase 0.0) and Min II, respectively.

The DC program of the W-D code produces the error estimates computed from the covariance matrix evaluated around best-fitting 
model light curve. But, it should be noted that the errors are unrealistically small because of the strong correlations 
between parameters \citep{maceroni1997,southworth2011}. In order to estimate more reliable errors, we followed
the procedure described by \citet{southworth2011}. First of all, the LOAO data for each object were splitted into five subsets
and fitted with the W-D code individually. Then, we calculated the standard deviations for the adjusted parameters, which were
not divided by the square root of the number of estimates for each parameter. Those values are close to the formal errors 
yielded by the W-D code from all observations. The error estimates presented in Tables \ref{tab6} and \ref{tab7} are adopted from the larger of 
the two values. 

\subsection{J093010A}

The light curves of J093010A display sharp eclipses and two light maxima are displaced around the secondary eclipse. 
However, the maxima were shown at just before and after the primary eclipse in the light curves given by \citet{lohr2013}.
In low-mass detached binaries, cool starspots produced by magnetic dynamo-related activity might be the reason for the 
seasonal light variations \citep[e.g., NSVS 02502726,][]{lee2013}. In order to obtain a unique binary solution, 
we simultaneously analyzed the RV and light curves in a manner similar to 
that for NSVS 02502726. The W-D code was applied to the light curves normalized to unit light at phase 0.55. 

\begin{figure}
 \includegraphics[]{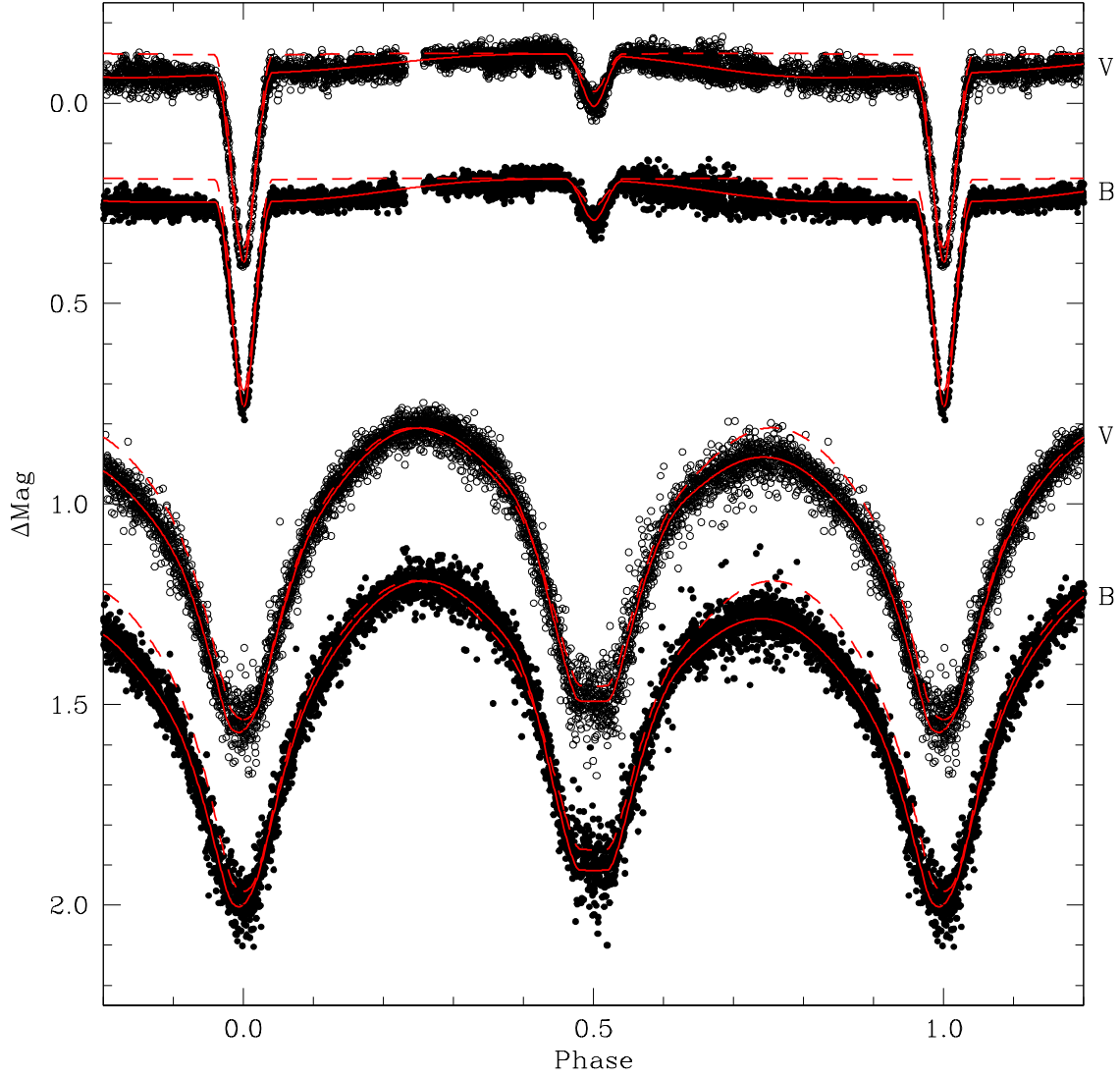}
 \caption{$BV$ light curves of J093010. The solid and dashed curves represent the spotted and unspotted solutions, respectively, obtained with the model parameters listed 
 in Table \ref{tab6} and \ref{tab7}. The upper and lower parts represent the light curves of J093010A and J093010B, respectively.
 \label{Fig2}}
\end{figure}

Our analyses have been carried out through two stages: For the first stage, the RV and light curves were solved without spots. 
The result for this unspotted model is listed in the columns (2)$-$(3) of Table \ref{tab6} and plotted as dashed curves in Figure \ref{Fig2}, 
where the computed light curves do not fit well with the LOAO observations. In the second stage, the spot parameters were 
included as additional parameters. There is, at present, no way to know which spot model is more efficient in creating 
the light variations other than by complete light-curve analysis. Firstly, we tested a single spot on either binary component 
of J093010A. The results improve the light-curve fitting partly but large discrepancies still remain. This may indicate 
the existence of a (or some) further spot(s). Final results are given in the columns (4)$-$(5) of Table \ref{tab6}. 
The synthetic light curves from our model are plotted as solid curves in the upper part of Figure \ref{Fig2}, while 
the synthetic RV curves are plotted in Figure \ref{Fig5}. As shown in the figures, the two-spot model with one cool spot 
on each component describes the conspicuous light variations of the system quite well. Separate trials for 
the other spot configurations were not so successful as was our model and even three-spot models do not give a better fit 
than do the two-spot model. In all the procedures, we included an orbital eccentricity for the binary orbit as a free parameter, 
but found that the parameter remained zero within its error.

The light and RV solutions represent the eclipsing pair of J093010A as a well-detached binary: the primary and 
secondary components fill 41.8\% and 46.7\% of their inner Roche lobes, respectively. This implies that there is 
no mass transfer between the two component stars. The third component has a light contribution of about 30\% and was detected 
at spectroscopic spectra. As a result, we obtained the absolute dimensions for the eclipsing pair given in the columns (2)$-$(3) 
of Table 9. The radii are the mean volume radii calculated from the tables of \citet{mochnacki1984} and the luminosity ($L$) 
was computed by adopting $T_{\rm eff}$$_\odot$ = 5,780 K and $M_{\rm bol}$$_\odot$ = +4.73 for solar values. 
The bolometric corrections (BCs) were obtained from the relation between $\log T_{\rm eff}$ and BC given by \citet{torres2010}. 
With the brightest apparent visual magnitude of $V$ = 9.70 $\pm$ 0.02, our computed light ratio, and 
the interstellar reddening of $A_{\rm V}$ = 0.03, we calculated the distance to the system to be 66 pc.

\subsection{J093010B}

The light curves of J093010B are the typical shape of a short period overcontact binary and show a flat bottom at 
the secondary minimum, indicating that this system belongs to the A-type of W UMa stars. Further, our observations display 
the O'Connell effect that Max I is brighter than Max II by 0.095 and 0.074 mag for the $B$ and $V$ bandpasses, respectively. 
These phenomena have been commonly reported for light curves of W UMa-type binaries and can be explained by the spot activity 
from magnetic dynamo and/or impact from mass transfer between components \citep[e.g., TU Boo,][]{lee2007}. Since no spectra 
of J093010B were detected, its light curves were analyzed in a manner similar to that for V432 Per \citep{lee2008} using 
the so-called $q$-search procedure.

To derive the mass ratio of the system, we calculated an intensive $q$-search for a series of models over the range of $q \le$ 1.0 
for two cases: with and without spot model. From the $q$-search results, the optimal solution is around $q$ = 0.45, which was 
treated as an adjustable parameter in all subsequent syntheses deriving binary parameters. As in the case of J093010A, 
various spot models were examined to explain the light discrepancy. The results are listed in Table \ref{tab7} and displayed in 
the lower part of Figure \ref{Fig2} with the dashed and solid curves. In the figure, the asymmetrical light curves are best 
modeled using a two-spot model with both a cool spot on the primary star and a hot spot on the secondary. The cool spot may have 
been formed by a magnetic dynamo, whereas the hot spot may have been produced by impact from mass transfer between the components. 

Our light-curve solutions indicate that J093010B is an overcontact binary with a mass ratio of $q$ = 0.468, an orbital inclination 
of $i$ = 87$^\circ$.0, a fill-out factor of $f$ = 21\%, and a third light source of $l_3$ = 4$\sim$5\%. Here, 
$f$ = ($\Omega_{\rm in}$--$\Omega$)/($\Omega_{\rm in}$--$\Omega_{\rm out}$), where the potentials 
$\Omega_{\rm in}$ and $\Omega_{\rm out}$ define the inner and outer critical surfaces in Roche geometry and $\Omega$ is 
the potential corresponding to the surface of the overcontact binary. The effective temperature of the primary component was 
estimated at the secondary eclipse phase where only the primary is visible. It corresponds to a normal main-sequence star with 
a spectral type of about K3. We estimated the absolute dimensions for J093010B using our photometric solution and 
\citet{harmanec1988}'s relation between the spectral type and stellar mass. The results are given in the columns (4)$-$(5) 
of Table \ref{tab8}, where the distance to the system of about 77 pc was calculated using the brightest apparent visual magnitude of 
$V$ = 10.62 $\pm$ 0.02 and the computed light ratio at phase 0.25.

\section{Summary and Discussion}
We performed photometric and spectroscopic follow-up observations for J093010 showing dual types of light curves.
Our PSF photometry revealed the brighter object J093010A to be an Algol-type eclipsing binary and the fainter J093010B to be a W UMa-type eclipsing binary. 
Furthermore, the third component of J093010A was detected via spectroscopic observations and its light contribution was calculated to be about 30\%.
However, we could not detect any line in the spectra of J093010B probably due to relatively long exposure time to its orbital period. 
The RVs of J093010A were measured and its absolute dimensions were derived from the simultaneous analysis of light curves and RVs.
The masses and radii of the components were found with an accuracy of $\sim$2\% and $\sim$1\%, respectively.

\begin{figure}
 \includegraphics[]{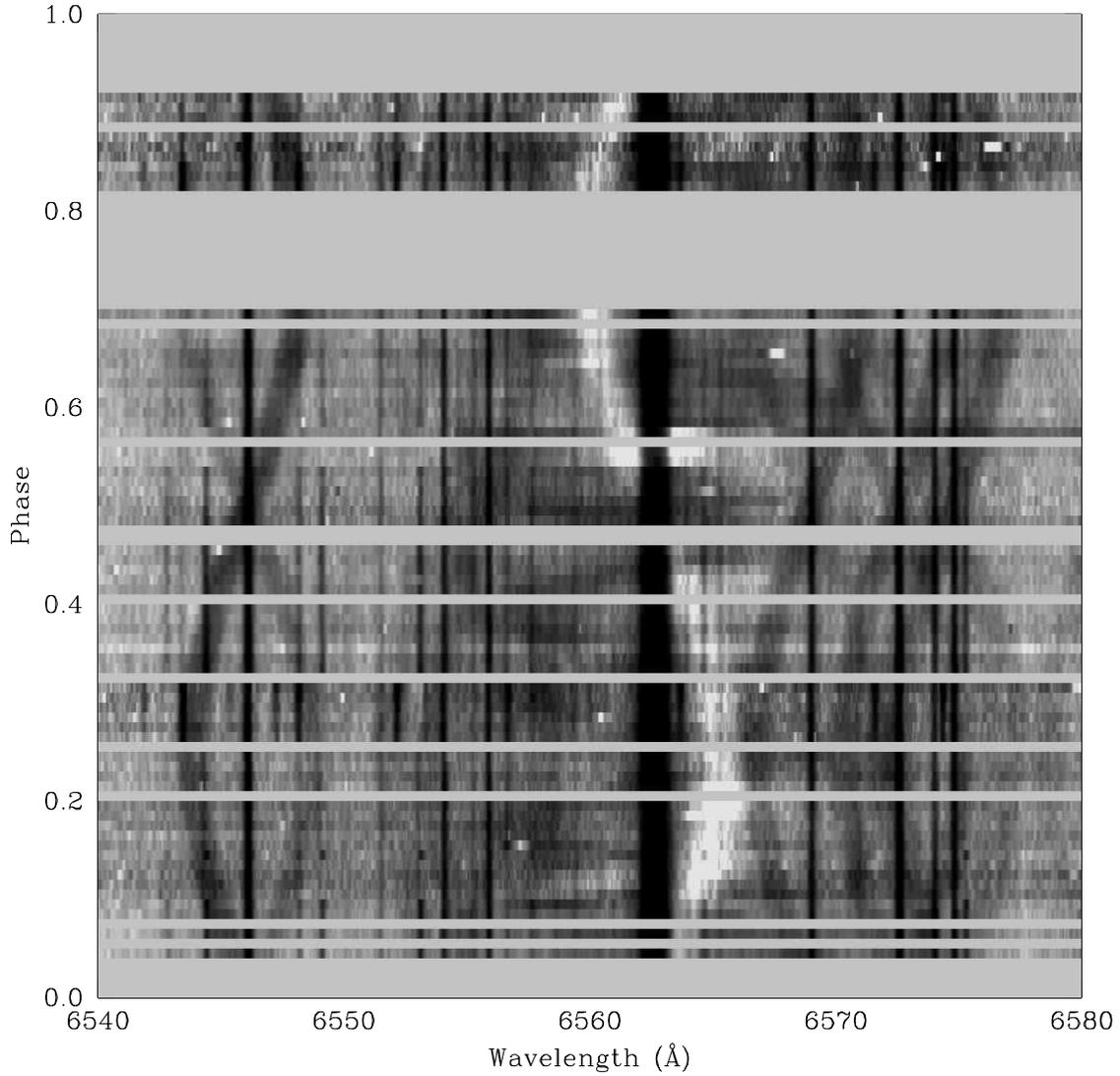}
 \caption{Variability of the spectra for J093010 based on the orbital phase for J093010A. One stationary and two variable components are displayed. H$\alpha$ 6563 \AA\, emissions are shown through the secondary orbit but any absorption or emission for the primary component is absent.
\label{Fig3}}
\end{figure}

\begin{figure}
 \includegraphics[]{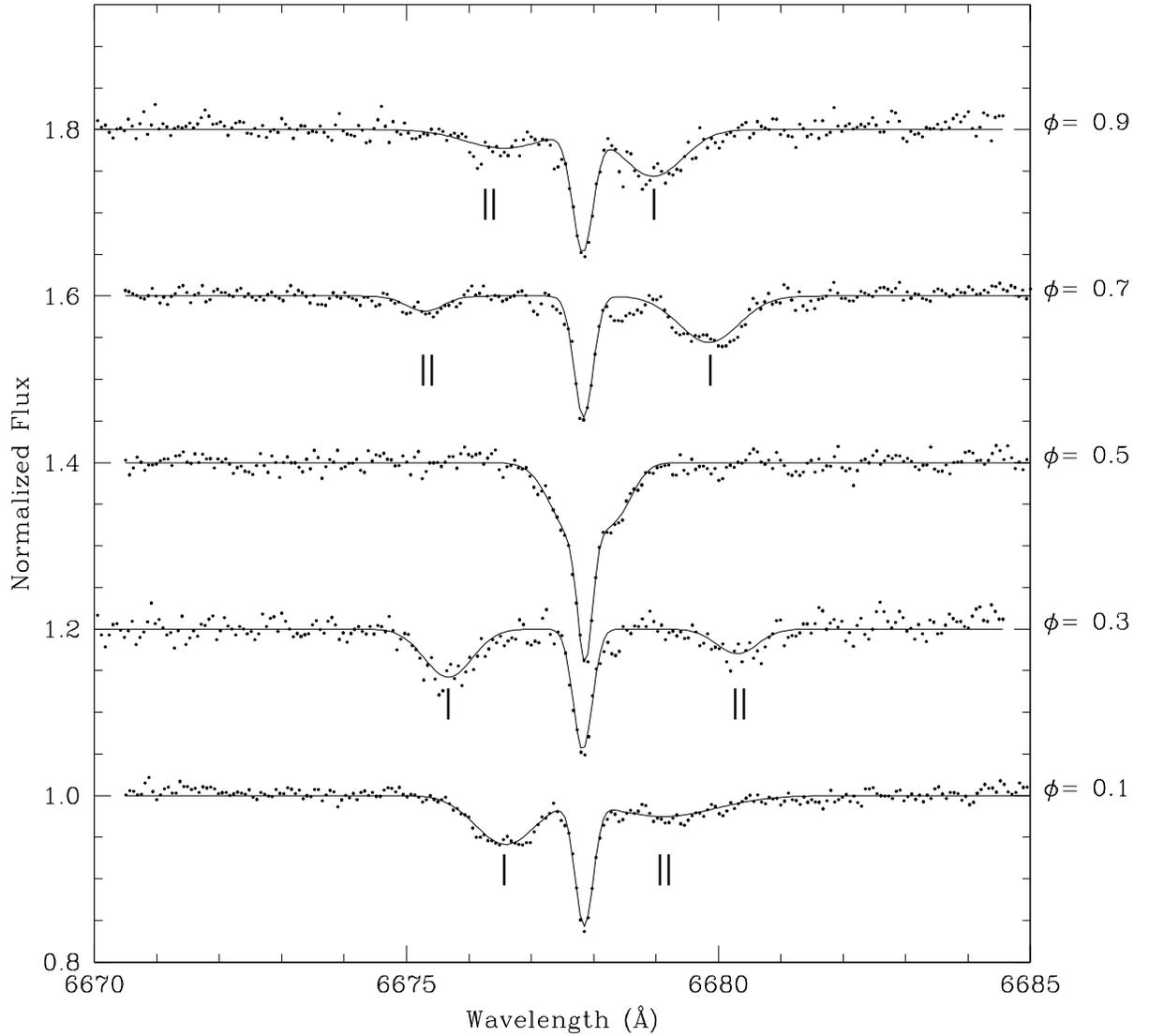}
 \caption{Sample spectra for 6678 \AA\, of Fe ${\rm I}$ lines to determine the RVs of three components with line-profile fitting using gaussian functions. 
 Dots and solid lines represent the observations and the fits for selected phases ($\phi$ = 0.1, 0.3, 0.5, 0.7, and 0.9), respectively.
 `$|$' for primary and `$||$' for secondary components are marked.
 Stationary absorption lines display the third component.
\label{Fig4}}
\end{figure}

\begin{figure}
 \includegraphics[]{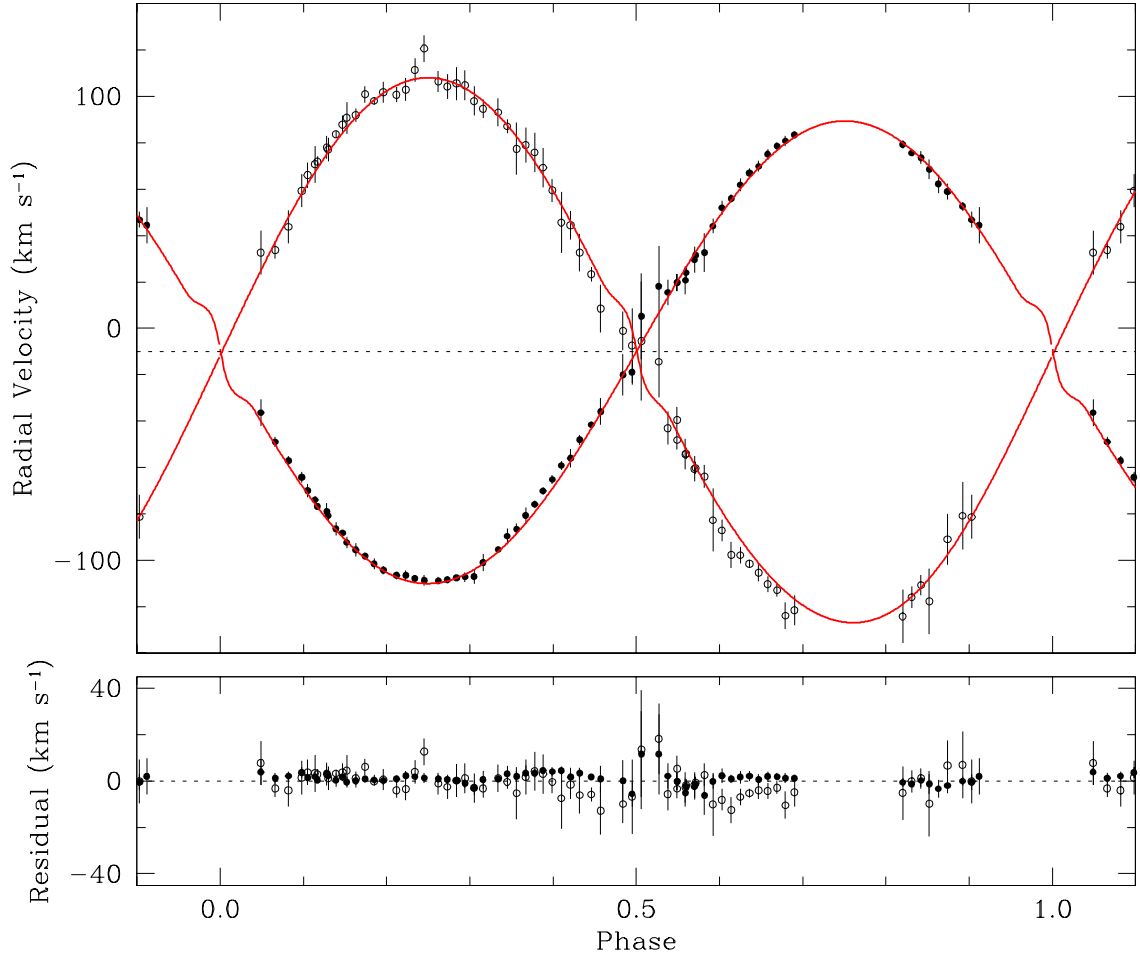}
 \caption{RV curves of J093010A. 
 Filled and open circles represent the RV measurements for primary and secondary components, respectively. 
 In the upper panel, the solid curves denote the result from consistent light and velocity curve analysis 
 and the dotted line refers to the system velocity of $-$10.1 km s$^{-1}$.
 Lower panel displays the RV residuals.
\label{Fig5}}

\end{figure}

\begin{figure}
 \includegraphics[]{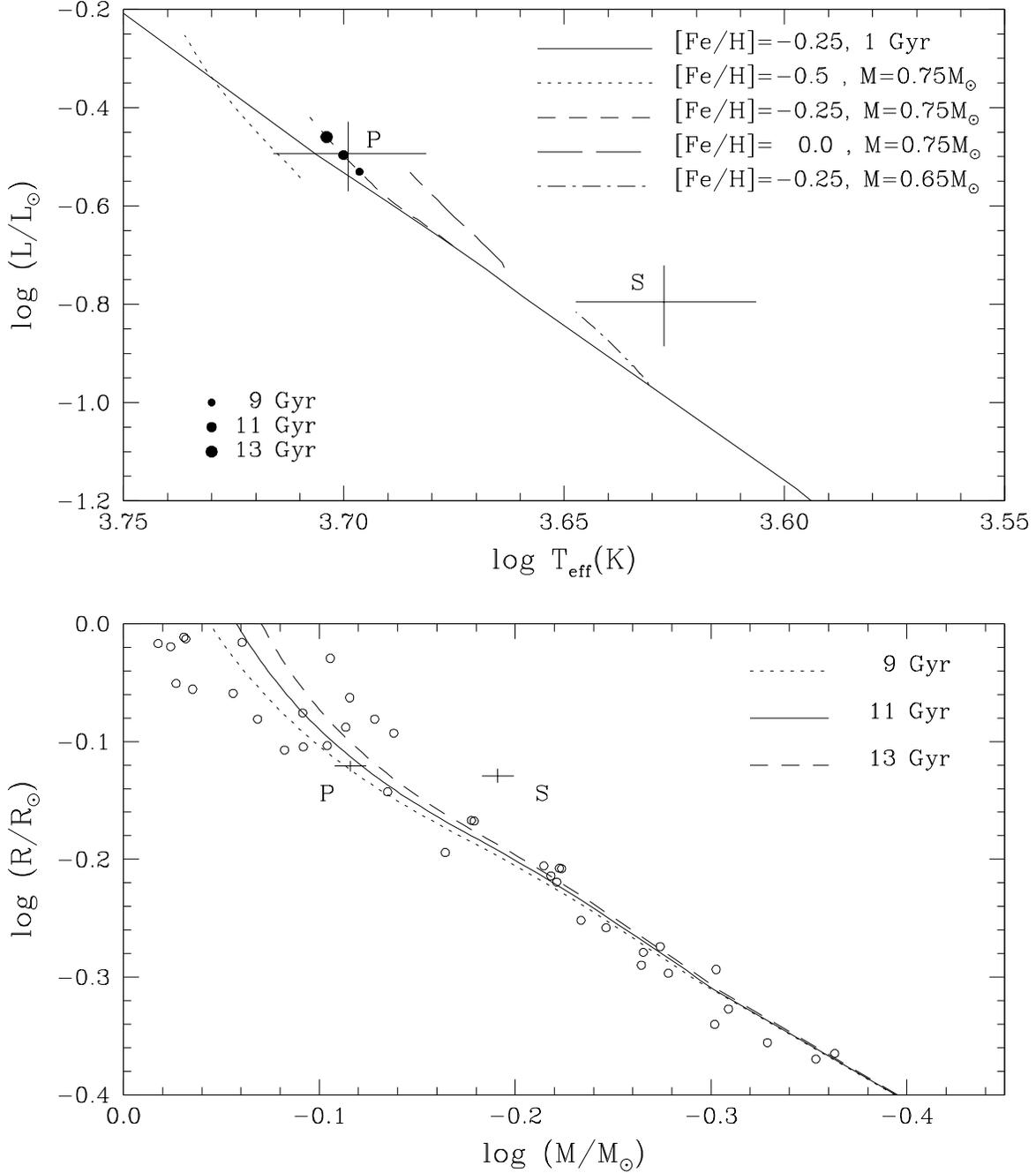}
 \caption{Comparisons of our absolute dimensions for J093010A with DSEP models and well-studied samples. In the upper panel, a solid line represents the isochrone of 1 Gyr and other lines show the evolutionary tracks from 1 to 15 Gyr having different metallicities and masses. And the positions of 9, 11, and 13 Gyr are marked with filled circles on the evolutionary track of [Fe/H] = $-$0.25 and M = 0.75 M$_\odot$. In the lower panel, isochrones of 9, 11 and 13 Gyr with [Fe/H] = $-$0.25 are represented with different line types. Open circles represent the well-studied detached binary systems from DEBCat. The primary star is comparable to the normal main-sequence star, but the secondary one has a larger size and lower temperature than those expected in the models and other low-mass stars.
\label{Fig6}}
\end{figure}

We compared our accurate physical parameters of J093010A with Dartmouth Stellar Evolution Program (DSEP) models \citep{dotter2008}. 
First, several evolutionary tracks with different metallicities were compared to the position of the primary component in the HR diagram.
In the upper panel of Figure \ref{Fig6}, three evolutionary tracks of  M = 0.75 M$_\odot$ are displayed on the HR diagram. 
The primary component is on the track of [Fe/H] = $-$0.25 (dashed line). Assuming the same metal abundances of two components, 
the secondary star (M = 0.644 M$_\odot$) is located on the right-hand side of the track for [Fe/H] = $-$0.25 and M = 0.65 M$_\odot$. 
We marked the positions of 9, 11, and 13 Gyr with filled circles in the panel.
We also compared our parameters of J093010A with the empirical mass-radius relation from well-studied detached binaries (from DEBCat at http://www.astro.keele.ac.uk/jkt/debcat/).
While the primary component is on the typical trend of main-sequence stars and on the isochrone of Age = 11 Gyr and [Fe/H] = $-$0.25, 
the secondary component has a larger radius than typical ones and is on quite above the isochrone (see lower panel of Figure \ref{Fig6}).
From the locations on the figure, the secondary component of J093010A has a relatively larger size and lower surface temperature than 
those expected from the model and empirical data.
The main reason for these disagreements might be associated with the magnetic activities that caused the cool spot on the surface and H$\alpha$ emissions \citep{torres2013}. 
Chromospheric active stars showing H$\alpha$ emission (see Figure \ref{Fig3}) have 
a lower T$_{\rm{eff}}$ and larger radius relative to both the theoretical isochrones and the non-active stars \citep{stassun2012}. 
All these magnetic activities and changes of the light maximum are well-known characteristics of RS CVn type systems. 
The count rate in the X-ray band (0.1 - 2.4 keV) for J093010 is 0.433 counts s$^{-1}$ \citep{voges1999} and 
the corresponding X-ray luminosity is calculated to be $\sim$10$^{30}$ ergs s$^{-1}$, assuming the distance of 70 pc. 
This value is in the typical range for the RS CVn systems \citep{dempsey1993}.

\citet{lohr2013} suggested J093010 to be a quadruple system that consists of two binaries with a common proper motion. 
As noted in the Tycho Double Star Catalogue \citep{fabricius2002}, it is the photocentre proper motion for a close double which was neither resolved in Tycho-2 nor measured individually.
We calculated the distances to each binary to be 66 $\pm$ 7 and 77 $\pm$ 9 pc, respectively.
The value for J093010B was estimated from an empirical relation between the temperature and stellar mass, 
and hence the error may be larger than the presented one. 
Two eclipsing binaries have almost same inclinations, similar masses and spectral types, which 
seem to be located at the same distance and to be gravitationally bound.
The distance to the third component of J093010A, assuming a main-sequence star, is also comparable to those of the two binaries. 
Moreover, the system velocity of the Algol-type binary is almost consistent with the RVs of the stationary component. This means that the third body is possibly bound to the eclipsing pair.
In conclusion, we suggest that J093010 is a possible quintuple system with a hierarchical structure of a triple A(ab)c and a binary B(ab).
Assuming that the distance to the system is 70 pc and the mass of the third component of J093010A is 0.7 M$_\odot$, the separation and meta-orbital period between the two systems (J093010A and B) are calculated to be $\sim$130 AU and $\sim$850 years, respectively.
While the RVs for the third component of J093010A seem to have slightly decreased by about 1 km s$^{-1}$ during our observation interval of $\sim$100 days, 
the corresponding increment ($\sim$0.5 km s$^{-1}$) of system velocities for J093010A(ab) was too small to measure.
High-precision and long-term observations are required to detect the variations.
In addition, the spectroscopic observations for J093010B will be helpful to obtain more accurate physical parameters and distance to the binary.



\clearpage
\begin{deluxetable}{cccc}
\tabletypesize{\scriptsize}
\tablewidth{0pt}
\tablecaption{Spectroscopic and photometric observation logs
\label{tab0}}
\tablehead{\multicolumn{4}{c}{Spectroscopy}\\
\colhead{DATE (UT)}   & \colhead{fiber ($\mu m$)} & \colhead{exp. time (sec.)} & \colhead{weather}
}
\startdata  
2012.11.18 &   300 &   1800  &    clear        \\   
2012.11.20 &   300 &   1200  &    clear        \\   
2012.11.21 &   300 &   1200  &    partly cloudy\\   
2012.12.09 &   300 &   1200  &    partly cloudy\\   
2013.02.22 &   300 &   1200  &    clear        \\   
2013.03.21 &   300 &   1200  &    clear        \\   
2013.03.22 &300, 80&   1200  &    partly cloudy\\   
2013.03.23 &   300 &   1200  &    cloudy        \\   \cline{1-4}
\multicolumn{4}{c}{Photometry}\\
DATE (UT)  & filter&   seeing (arcsec)&  weather        \\   \cline{1-4}
2012.12.06 &   $BV$  &   2.0 &   partly cloudy \\ 
2012.12.07 &   $BV$  &   1.6 &   partly cloudy \\ 
2012.12.08 &   $BV$  &   1.4 &   clear         \\ 
2012.12.09 &   $BV$  &   1.4 &   partly cloudy \\ 
2012.12.10 &   $BV$  &   2.4 &   partly cloudy \\ 
2012.12.27 &   $BV$  &   1.9 &   clear, full moon \\ 
2013.01.05 &   $BV$  &   1.5 &   clear         \\ 
2013.01.06 &   $BV$  &   1.7 &   partly cloudy \\ 
2013.01.07 &   $BV$  &   1.6 &   partly cloudy \\ 
2013.01.08 &   $BV$  &   1.7 &   partly cloudy \\ 
2013.01.09 &   $BV$  &   2.1 &   clear         \\ 
2013.01.10 &   $BV$  &   1.7 &   partly cloudy \\ 
2013.02.16 &   $BV$  &   2.1 &   clear         \\ 
2013.02.23 &   $BV$  &   1.8 &   clear         \\ 
2013.02.24 &   $BV$  &   1.9 &   clear, full moon \\ 
                                         
\enddata                                      
\end{deluxetable}

\begin{deluxetable}{cccccc}
\tabletypesize{\scriptsize}
\tablewidth{0pt}
\tablecaption{CCD photometric data of J093010A and B
\label{tab1}}
\tablehead{
\colhead{HJD}  & \multicolumn{2}{c}{$\Delta B$}  &  \colhead{HJD}  & \multicolumn{2}{c}{$\Delta V$}  \\  \cline{2-3} \cline{5-6}\\[-2.0ex]
\colhead{} & \colhead{J093010A} & \colhead{J093010B} & \colhead{} & \colhead{J093010A} & \colhead{J093010B} 
}
\startdata  
2456267.76993  &   0.218  &   1.196 &	2456267.77023  &  $-$0.088  &  0.819 \\
2456267.77057  &   0.200  &   1.248 &	2456267.77088  &  $-$0.115  &  0.855 \\
2456267.77122  &   0.223  &   1.194 &	2456267.77153  &  $-$0.054  &  0.840 \\
2456267.77187  &   0.246  &   1.156 &	2456267.77218  &  $-$0.085  &  0.810 \\
2456267.77252  &   0.224  &   1.212 &	2456267.77283  &  $-$0.098  &  0.917 \\
2456267.77317  &   0.237  &   1.217 &	2456267.77347  &  $-$0.079  &  0.837 \\
2456267.77382  &   0.216  &   1.240 &	2456267.77412  &  $-$0.107  &  0.889 \\
2456267.77446  &   0.268  &   1.164 &	2456267.77477  &  $-$0.051  &  0.818 \\
2456267.77510  &   0.196  &   1.268 &	2456267.77541  &  $-$0.066  &  0.866 \\
2456267.77575  &   0.198  &   1.271 &	2456267.77606  &  $-$0.078  &  0.885 \\
\enddata
\tablecomments{This table  is published in its entirety in the electronic edition of the journal.
A portion is shown here for guidance regarding its form and content}
\end{deluxetable}

\begin{deluxetable}{lccc}
\tabletypesize{\scriptsize} 
\tablewidth{0pt}
\tablecaption{Minimum times for J093010A (Algol-type)
\label{tab2}}
\tablehead{
\colhead{HJD}  & \colhead{Error}  & \colhead{Min}  &  \colhead{Filter}
}
\startdata  
2456267.80132 & $\pm$0.00017 &  II  & $BV$\\
2456297.82710 & $\pm$0.00019 &  II  & $BV$\\
2456303.05239 & $\pm$0.00015 &  II  & $BV$\\
2456346.78443 & $\pm$0.00015 &  I   & $BV$\\
\enddata
\end{deluxetable}

\begin{deluxetable}{lccc}
\tabletypesize{\scriptsize}
\tablewidth{0pt}
\tablecaption{Minimum times for J093010B (W UMa-type)
\label{tab3}}
\tablehead{
\colhead{HJD} & \colhead{Error}  & \colhead{Min}  &  \colhead{Filter}
}
\startdata
2456267.81536 & $\pm$0.00023 &  II  & $BV$\\
2456267.92899 & $\pm$0.00030 &  I   & $BV$\\
2456268.83924 & $\pm$0.00051 &  I   & $BV$\\
2456269.86573 & $\pm$0.00017 &  II  & $BV$\\
2456269.97770 & $\pm$0.00015 &  I   & $BV$\\
2456270.88880 & $\pm$0.00017 &  I   & $BV$\\
2456271.00394 & $\pm$0.00015 &  II  & $BV$\\
2456288.87879 & $\pm$0.00020 &  I   & $BV$\\
2456297.87410 & $\pm$0.00019 &  II  & $BV$\\
2456297.98655 & $\pm$0.00024 &  I   & $BV$\\
2456298.89837 & $\pm$0.00016 &  I   & $BV$\\
2456299.01309 & $\pm$0.00026 &  II  & $BV$\\
2456299.92307 & $\pm$0.00025 &  II  & $BV$\\
2456300.03608 & $\pm$0.00016 &  I   & $BV$\\
2456300.94721 & $\pm$0.00019 &  I   & $BV$\\
2456301.85768 & $\pm$0.00030 &  I   & $BV$\\
2456301.97328 & $\pm$0.00059 &  II  & $BV$\\
2456302.88386 & $\pm$0.00061 &  II  & $BV$\\
2456302.99612 & $\pm$0.00031 &  I   & $BV$\\
2456340.00143 & $\pm$0.00050 &  II  & $BV$\\
2456347.85636 & $\pm$0.00018 &  I   & $BV$\\
\enddata
\end{deluxetable}

\clearpage
\begin{deluxetable}{lrrr}
\tabletypesize{\scriptsize}
\tablewidth{0pt}
\tablecaption{Radial velocities for J093010A
\label{tab4}}
\tablehead{
\colhead{HJD} & \colhead{RV$_{\rm{Pri.}}$} & \colhead{RV$_{\rm{Sec.}}$} & \colhead{RV$_{\rm{3rd.}}$} \\ \cline{2-4} \\[-2.0ex]
\colhead{}    & \colhead{km s$^{-1}$}      & \colhead{km s$^{-1}$} & \colhead{km s$^{-1}$} 
}
\startdata 
2456250.24230 &    $-$36.33  $\pm$     5.66 &       32.68 $\pm$      9.45 &     $-$6.72  $\pm$     0.57 \\
2456250.26347 &    $-$48.95  $\pm$     2.27 &       33.78 $\pm$      3.64 &     $-$6.54  $\pm$     0.73 \\
2456250.28460 &    $-$57.12  $\pm$     2.02 &       43.77 $\pm$      7.03 &     $-$6.25  $\pm$     0.74 \\
2456250.30574 &    $-$64.17  $\pm$     2.28 &       59.32 $\pm$      7.20 &     $-$7.55  $\pm$     1.60 \\
2456250.32687 &    $-$73.90  $\pm$     1.66 &       70.71 $\pm$      7.95 &     $-$6.48  $\pm$     0.44 \\
2456250.34801 &    $-$80.75  $\pm$     1.99 &       77.09 $\pm$      5.12 &     $-$6.45  $\pm$     0.49 \\
2456250.36916 &    $-$88.12  $\pm$     1.28 &       87.85 $\pm$      3.76 &     $-$6.18  $\pm$     0.61 \\
2456252.11499 &    $-$20.02  $\pm$     8.98 &     $-$1.07 $\pm$      8.31 &     $-$6.75  $\pm$     0.53 \\
2456252.12917 &    $-$18.87  $\pm$     5.35 &     $-$7.40 $\pm$     15.96 &     $-$6.66  $\pm$     0.80 \\
2456252.14336 &        5.24  $\pm$    18.50 &     $-$5.45 $\pm$     25.59 &     $-$6.80  $\pm$     0.56 \\
2456252.17171 &       18.15  $\pm$    17.32 &    $-$14.41 $\pm$     15.23 &     $-$6.97  $\pm$     0.66 \\
2456252.18588 &       15.48  $\pm$     5.39 &    $-$42.99 $\pm$      7.10 &     $-$6.46  $\pm$     0.62 \\
2456252.20005 &       19.76  $\pm$     3.70 &    $-$48.12 $\pm$      4.21 &     $-$6.43  $\pm$     0.45 \\
2456252.21422 &       23.99  $\pm$     2.66 &    $-$53.99 $\pm$      2.45 &     $-$6.75  $\pm$     0.52 \\
2456252.22840 &       31.50  $\pm$     1.69 &    $-$60.15 $\pm$      2.62 &     $-$7.05  $\pm$     0.84 \\
2456252.24258 &       32.70  $\pm$     8.31 &    $-$63.80 $\pm$      4.90 &     $-$7.29  $\pm$     0.80 \\
2456252.25676 &       44.03  $\pm$     3.17 &    $-$82.62 $\pm$     13.53 &     $-$7.09  $\pm$     0.63 \\
2456252.27094 &       52.00  $\pm$     3.11 &    $-$87.04 $\pm$      4.67 &     $-$7.09  $\pm$     0.55 \\
2456252.28511 &       55.94  $\pm$     1.85 &    $-$97.56 $\pm$      5.65 &     $-$7.15  $\pm$     0.73 \\
2456252.29928 &       61.80  $\pm$     2.66 &    $-$97.73 $\pm$      3.62 &     $-$7.00  $\pm$     0.53 \\
2456252.31345 &       66.93  $\pm$     2.13 &   $-$101.51 $\pm$      1.86 &     $-$7.09  $\pm$     0.43 \\
2456252.32765 &       69.81  $\pm$     2.34 &   $-$105.28 $\pm$      3.72 &     $-$7.04  $\pm$     0.49 \\
2456252.34182 &       75.06  $\pm$     2.20 &   $-$110.14 $\pm$      3.47 &     $-$7.05  $\pm$     0.33 \\
2456252.35599 &       78.53  $\pm$     1.55 &   $-$112.83 $\pm$      2.68 &     $-$6.89  $\pm$     0.65 \\
2456252.37016 &       80.73  $\pm$     2.19 &   $-$123.78 $\pm$      5.88 &     $-$6.87  $\pm$     0.58 \\
2456252.38433 &       83.43  $\pm$     1.45 &   $-$121.45 $\pm$      6.35 &     $-$6.87  $\pm$     0.43 \\
2456253.22487 &    $-$95.27  $\pm$     1.22 &       93.08 $\pm$      5.94 &     $-$6.75  $\pm$     0.75 \\
2456253.23905 &    $-$89.56  $\pm$     3.46 &       87.11 $\pm$      3.01 &     $-$6.76  $\pm$     0.75 \\
2456253.25325 &    $-$86.52  $\pm$     2.39 &       77.41 $\pm$     11.29 &     $-$7.02  $\pm$     0.63 \\
2456253.26751 &    $-$80.55  $\pm$     3.49 &       79.03 $\pm$      7.60 &     $-$6.78  $\pm$     0.52 \\
2456253.28169 &    $-$75.83  $\pm$     1.77 &       75.85 $\pm$      8.39 &     $-$6.77  $\pm$     0.61 \\
2456253.29587 &    $-$70.03  $\pm$     1.57 &       69.22 $\pm$      8.49 &     $-$6.74  $\pm$     0.69 \\
2456253.31005 &    $-$65.10  $\pm$     1.94 &       59.46 $\pm$      4.88 &     $-$6.88  $\pm$     0.70 \\
2456253.32423 &    $-$59.08  $\pm$     1.88 &       45.63 $\pm$     13.16 &     $-$6.51  $\pm$     0.64 \\
2456253.33840 &    $-$55.94  $\pm$     3.94 &       44.44 $\pm$      6.28 &     $-$6.96  $\pm$     0.77 \\
2456253.35258 &    $-$48.01  $\pm$     2.21 &       32.64 $\pm$      8.08 &     $-$6.91  $\pm$     0.44 \\
2456253.37069 &    $-$41.55  $\pm$     1.26 &       23.36 $\pm$      3.10 &     $-$7.38  $\pm$     0.76 \\
2456253.38487 &    $-$35.87  $\pm$     5.77 &        8.59 $\pm$     10.33 &     $-$7.37  $\pm$     0.96 \\
2456271.20294 &    $-$69.92  $\pm$     2.75 &       66.03 $\pm$      5.43 &     $-$6.43  $\pm$     0.64 \\
2456271.21913 &    $-$76.75  $\pm$     1.02 &       71.86 $\pm$      2.20 &     $-$6.40  $\pm$     0.62 \\
2456271.23332 &    $-$78.93  $\pm$     3.53 &       77.96 $\pm$      4.90 &     $-$6.61  $\pm$     0.65 \\
2456271.24750 &    $-$86.38  $\pm$     2.84 &       83.69 $\pm$      1.58 &     $-$6.29  $\pm$     0.58 \\
2456271.26433 &    $-$92.23  $\pm$     2.41 &       90.71 $\pm$      6.80 &     $-$6.65  $\pm$     0.82 \\
2456271.27851 &    $-$95.46  $\pm$     2.80 &       91.92 $\pm$      2.85 &     $-$6.34  $\pm$     0.60 \\
2456271.29269 &    $-$98.09  $\pm$     1.34 &      100.84 $\pm$      3.60 &     $-$6.52  $\pm$     0.53 \\
2456271.30687 &   $-$101.51  $\pm$     2.31 &       98.04 $\pm$      1.57 &     $-$6.51  $\pm$     0.75 \\
2456271.32105 &   $-$104.19  $\pm$     1.86 &      101.81 $\pm$      4.53 &     $-$6.31  $\pm$     0.72 \\
2456271.34241 &   $-$106.41  $\pm$     1.64 &      100.71 $\pm$      3.23 &     $-$6.93  $\pm$     0.78 \\
2456271.35659 &   $-$106.41  $\pm$     2.15 &      102.85 $\pm$      4.93 &     $-$7.01  $\pm$     0.72 \\
2456271.37085 &   $-$107.73  $\pm$     0.89 &      111.33 $\pm$      5.17 &     $-$7.34  $\pm$     0.63 \\
2456271.38503 &   $-$108.64  $\pm$     2.27 &      120.57 $\pm$      5.82 &     $-$7.50  $\pm$     0.92 \\
2456346.19552 &       19.75  $\pm$     3.83 &    $-$39.47 $\pm$      5.54 &     $-$7.29  $\pm$     0.48 \\
2456346.20970 &       20.74  $\pm$     6.05 &    $-$54.28 $\pm$      6.59 &     $-$7.57  $\pm$     0.51 \\
2456346.22388 &       29.62  $\pm$     5.49 &    $-$60.51 $\pm$      5.52 &     $-$7.85  $\pm$     0.70 \\
2456373.23689 &   $-$108.76  $\pm$     1.63 &      106.42 $\pm$      4.60 &     $-$7.60  $\pm$     0.64 \\
2456373.25107 &   $-$108.27  $\pm$     1.50 &      104.16 $\pm$      5.29 &     $-$7.43  $\pm$     0.57 \\
2456373.26524 &   $-$107.54  $\pm$     1.00 &      105.51 $\pm$      7.12 &     $-$7.50  $\pm$     0.83 \\
2456373.27940 &   $-$107.08  $\pm$     2.06 &      104.71 $\pm$      6.37 &     $-$7.25  $\pm$     0.76 \\
2456373.29358 &   $-$106.99  $\pm$     3.05 &       97.96 $\pm$      6.34 &     $-$7.66  $\pm$     0.43 \\
2456373.30774 &   $-$100.89  $\pm$     3.76 &       94.66 $\pm$      3.98 &     $-$7.26  $\pm$     0.63 \\
2456374.00695 &       68.53  $\pm$     4.26 &   $-$117.63 $\pm$     14.09 &     $-$8.33  $\pm$     0.75 \\
2456374.02114 &       62.23  $\pm$     3.91 &      \dots                  &     $-$9.67  $\pm$     1.32 \\
2456374.03531 &       58.95  $\pm$     3.39 &    $-$90.88 $\pm$     10.86 &     $-$7.50  $\pm$     1.24 \\
2456374.05973 &       52.63  $\pm$     1.93 &    $-$80.73 $\pm$     14.52 &     $-$8.12  $\pm$     1.00 \\
2456374.07393 &       46.76  $\pm$     3.43 &    $-$81.28 $\pm$      9.47 &     $-$7.87  $\pm$     0.87 \\
2456374.08555 &       44.43  $\pm$     7.82 &      \dots                  &     $-$8.56  $\pm$     0.96 \\
2456375.27153 &       79.29  $\pm$     1.87 &   $-$124.11 $\pm$     11.51 &     $-$7.80  $\pm$     0.94 \\
2456375.28570 &       75.57  $\pm$     1.02 &   $-$115.81 $\pm$      4.71 &     $-$7.89  $\pm$     0.65 \\
2456375.29988 &       73.57  $\pm$     2.82 &   $-$110.61 $\pm$      4.38 &     $-$7.72  $\pm$     0.78 \\
                                                
\enddata                                          
\end{deluxetable}

\begin{deluxetable}{lccccc}
\tabletypesize{\scriptsize} 
\tablewidth{0pt}
\tablecaption{Standard magnitudes and color indices for J093010AB
\label{tab5}}
\tablehead{
\colhead{Phase} & \multicolumn{2}{c}{J093010A} && \multicolumn{2}{c}{J093010B}  \\ 
[1.0mm] \cline{2-3} \cline{5-6} \\[-2.0ex]
\colhead{} & \colhead{$V$} & \colhead{($B-V$)} && \colhead{$V$} & \colhead{($B-V$)}
}
\startdata  
0.00       &   10.24(3)   &  1.03(4)           &&   11.35(3)    &   1.10(4)   \\
0.25       &   \dots      &  \dots             &&   10.62(2)    &   1.04(3)   \\
0.50       &   9.81(2)    &  0.97(2)           &&   11.29(2)    &   1.09(3)   \\
0.75       &   \dots      &  \dots             &&   10.69(2)    &   1.07(2)   \\
0.55       &   9.70(2)    &  0.99(2)           &&   \dots       &   \dots       \\
\enddata
\tablecomments{Normalized with UCAC4 catalogue.}
\end{deluxetable}

\begin{deluxetable}{lccccc}
\tabletypesize{\scriptsize}  
\tablewidth{0pt}
\tablecaption{RV and light curve parameters of J093010A \label{tab6}}
\tablehead{
\colhead{Parameter}                     & \multicolumn{2}{c}{Without Spot}           && \multicolumn{2}{c}{With Spot}                \\ [1.0mm] \cline{2-3} \cline{5-6} \\[-2.0ex]
                                        & \colhead{Primary} & \colhead{Secondary}    && \colhead{Primary} & \colhead{Secondary}
}                                                                                    
\startdata                                                                           
$T_0$ (HJD)                             & \multicolumn{2}{c}{2,456,346.78398(13)}    && \multicolumn{2}{c}{2,456,346.78382(11)}      \\
$P$ (day)                               & \multicolumn{2}{c}{1.3055432(33)}          && \multicolumn{2}{c}{1.3055394(26)}            \\
$\gamma$ (km s$^{-1}$)                  & \multicolumn{2}{c}{$-$10.06(77)}           && \multicolumn{2}{c}{$-$10.13(57)}             \\
$a$ (R$_\odot$)                         & \multicolumn{2}{c}{5.639(52)}              && \multicolumn{2}{c}{5.634(39)}                \\
$q$                                     & \multicolumn{2}{c}{0.841(15)}              && \multicolumn{2}{c}{0.841(10)}                \\
$i$ (deg)                               & \multicolumn{2}{c}{84.63(36)}              && \multicolumn{2}{c}{85.26(13)}                \\
$T$ (K)$\rm ^a$                         & 5000(200)         & 4040(200)              && 5000(200)         & 4240(200)                \\
$\Omega$                                & 8.96(10)          & 6.71(10)               && 8.349(51)         & 7.467(67)                \\
$\Omega_{\rm in}$                       & \multicolumn{2}{c}{3.487}                  && \multicolumn{2}{c}{3.486}                    \\
$X$, $Y$                                & 0.288,  0.408     & 0.142,  0.517          && 0.288,  0.408     & 0.254,  0.421            \\
$x_{B}$, $y_{B}$                        & 0.886, $-$0.041   & 0.935, $-$0.125        && 0.887, $-$0.042   & 1.197, $-$0.423          \\
$x_{V}$, $y_{V}$                        & 0.561,  0.273     & 0.614,  0.213          && 0.561,  0.273     & 0.799,  0.000            \\
$l$/($l_{1}$+$l_{2}$+$l_{3}$){$_{B}$}   & 0.534(27)         & 0.122                  && 0.553(11)         & 0.129                    \\
$l$/($l_{1}$+$l_{2}$+$l_{3}$){$_{V}$}   & 0.503(26)         & 0.166                  && 0.527(11)         & 0.164                    \\
{\it $l_{3B}$$\rm ^b$}                  & \multicolumn{2}{c}{0.344(30)}              && \multicolumn{2}{c}{0.318(12)}                \\
{\it $l_{3V}$$\rm ^b$}                  & \multicolumn{2}{c}{0.331(30)}              && \multicolumn{2}{c}{0.309(12)}                \\
$r$ (pole)                              & 0.1231(16)        & 0.1490(26)             && 0.1331(9)         & 0.1315(14)               \\
$r$ (point)                             & 0.1236(16)        & 0.1506(27)             && 0.1338(9)         & 0.1324(14)               \\
$r$ (side)                              & 0.1233(16)        & 0.1495(27)             && 0.1334(9)         & 0.1318(14)               \\
$r$ (back)                              & 0.1236(16)        & 0.1504(27)             && 0.1337(9)         & 0.1323(14)               \\
$r$ (volume)$\rm ^c$                    & 0.1233(16)        & 0.1497(27)             && 0.1344(9)         & 0.1319(14)               \\ [1.0mm]
\multicolumn{6}{l}{Spot parameters:}                                                                                                 \\ 
Colatitude (deg)                        & \dots             & \dots                  && 33.5              & 66.9                     \\
Longitude (deg)                         & \dots             & \dots                  && 334.8             & 257.8                    \\
Radius (deg)                            & \dots             & \dots                  && 37.6              & 45.0                     \\
$T$$\rm _{spot}$/$T$$\rm _{local}$      & \dots             & \dots                  && 0.939             & 0.913                    \\
$\Sigma W(O-C)^2$                       & \multicolumn{2}{c}{0.015}                  && \multicolumn{2}{c}{0.010}                
\enddata
\tablenotetext{a}{Errors were assigned according to the uncertainties in the color indices.}
\tablenotetext{b}{Value at 0.55 phase.}
\tablenotetext{c}{Mean volume radius.}
\end{deluxetable}

\clearpage
\begin{deluxetable}{lccccc}
\tabletypesize{\scriptsize}  
\tablewidth{0pt}
\tablecaption{Light curve parameters of J093010B \label{tab7}}
\tablehead{
\colhead{Parameter}                     & \multicolumn{2}{c}{Without Spot}           && \multicolumn{2}{c}{With Spot}                \\ [1.0mm] \cline{2-3} \cline{5-6} \\[-2.0ex]
                                        & \colhead{Primary} & \colhead{Secondary}    && \colhead{Primary} & \colhead{Secondary}
}
\startdata                                                                         
$T_0$ (HJD)                             & \multicolumn{2}{c}{2,456,267.92888(26)}    && \multicolumn{2}{c}{2,456,267.929523(38)}     \\
$P$ (day)                               & \multicolumn{2}{c}{0.22771622(35)}         && \multicolumn{2}{c}{0.22771581(25)}           \\
$q$                                     & \multicolumn{2}{c}{0.4680(68)}             && \multicolumn{2}{c}{0.4680(50)}               \\
$i$ (deg)                               & \multicolumn{2}{c}{86.93(90)}              && \multicolumn{2}{c}{87.09(68)}                \\
$T$ (K)$\rm ^a$                         & 4730(200)         & 4740(200)              && 4730(200)         & 4790(200)                \\
$\Omega$                                & 2.751(12)         & 2.751                  && 2.759(9)          & 2.759                    \\
$\Omega_{\rm in}$                       & \multicolumn{2}{c}{2.814}                  && \multicolumn{2}{c}{2.814}                    \\
$X$, $Y$                                & 0.306,  0.382     & 0.306,  0.382          && 0.306,  0.382     & 0.303,  0.386            \\
$x_{B}$, $y_{B}$                        & 1.039, $-$0.220   & 1.034, $-$0.214        && 1.039, $-$0.220   & 1.005, $-$0.180          \\
$x_{V}$, $y_{V}$                        & 0.689,  0.130     & 0.685,  0.134          && 0.689,  0.130     & 0.661,  0.161            \\
$l$/($l_{1}$+$l_{2}$+$l_{3}$){$_{B}$}   & 0.629(8)          & 0.325                  && 0.607(6)          & 0.341                    \\
$l$/($l_{1}$+$l_{2}$+$l_{3}$){$_{V}$}   & 0.632(8)          & 0.325                  && 0.616(6)          & 0.339                    \\
{\it $l_{3B}$$\rm ^b$}                  & \multicolumn{2}{c}{0.046(9)}               && \multicolumn{2}{c}{0.052(7)}                 \\
{\it $l_{3V}$$\rm ^b$}                  & \multicolumn{2}{c}{0.043(9)}               && \multicolumn{2}{c}{0.046(7)}                 \\
$r$ (pole)                              & 0.4308(24)        & 0.3061(34)             && 0.4294(18)        & 0.3046(25)               \\
$r$ (side)                              & 0.4608(33)        & 0.3212(42)             && 0.4589(24)        & 0.3193(31)               \\
$r$ (back)                              & 0.4930(47)        & 0.3621(77)             && 0.4905(34)        & 0.3590(57)               \\
$r$ (volume)                            & 0.4635(34)        & 0.3319(48)             && 0.4615(25)        & 0.3297(35)               \\ [1.0mm]
\multicolumn{6}{l}{Spot parameters:}                                                                                                 \\ 
Colatitude (deg)                        & \dots             & \dots                  && 89.5              & 90.7                     \\
Longitude (deg)                         & \dots             & \dots                  && 59.2              & 25.3                     \\
Radius (deg)                            & \dots             & \dots                  && 22.6              & 9.94                     \\
$T$$\rm _{spot}$/$T$$\rm _{local}$      & \dots             & \dots                  && 0.904             & 1.213                    \\
$\Sigma W(O-C)^2$                       & \multicolumn{2}{c}{0.019}                  && \multicolumn{2}{c}{0.013}                
\enddata
\tablenotetext{a}{Errors were assigned according to the uncertainties in the color indices.}
\tablenotetext{b}{Value at 0.25 phase.}
\end{deluxetable}

\clearpage
\begin{deluxetable}{lccccc}
\tablewidth{0pt}
\tablecaption{Absolute parameters for J093010 \label{tab8}}
\tablehead{
\colhead{Parameter}                     & \multicolumn{2}{c}{J093010A}               && \multicolumn{2}{c}{J093010B}                 \\ [1.0mm] \cline{2-3} \cline{5-6} \\[-2.0ex]
										                    & \colhead{Primary} & \colhead{Secondary}    && \colhead{Primary} & \colhead{Secondary}        
}                                                                                                                                      
\startdata                                                                                                                             
$M$ (M$_\odot$)                         & 0.766(14)         & 0.644(12)              && 0.738(60)       & 0.346(28)                  \\
$R$ (R$_\odot$)                         & 0.757(8)          & 0.743(10)              && 0.743(25)       & 0.531(19)                  \\
$\log$ $g$ (cgs)                        & 4.564(12)         & 4.505(14)              && 4.564(46)       & 4.527(47)                  \\
$\rho$ (g cm$^3)$                       & 2.49(9)           & 2.22(10)               && 2.54(33)        & 3.26(44)                   \\
$L$ (L$_\odot$)                         & 0.321(52)         & 0.160(30)              && 0.248(45)       & 0.133(24)                  \\
$M_{\rm bol}$ (mag)                     & +5.97(18)         & +6.72(21)              && +6.24(20)       & +6.92(20)                  \\
BC (mag)                                & $-$0.28(9)        & $-$0.83(21)            && $-$0.45(13)     & $-$0.41(12)                \\
$M_{\rm V}$ (mag)                       & +6.25(20)         & +7.55(29)              && +6.69(23)       & +7.33(23)                  \\
Distance (pc)                           & \multicolumn{2}{c}{66(7)}                  && \multicolumn{2}{c}{77(9)}                    \\
\enddata
\tablecomments{The values for J093010A are derived from simultaneous analysis of RVs and light curves, while those of J093010B are from light curve parameters and \citet{harmanec1988}'s relation.}
\end{deluxetable}

\clearpage


\newpage
\begin{thebibliography}{}

\bibitem[Ag{\"u}eros et al.(2009)]{agueros2009} Ag{\"u}eros, M.~A., Anderson, S.~F., Covey, K.~R., et al.\ 2009, \apjs, 181, 444 
\bibitem[Armstrong et al.(2012)]{armstrong2012} Armstrong, D., Pollacco, D., Watson, C.~A., et al.\ 2012, \aap, 545, L4 
\bibitem[Batten \& Hardie(1965)]{batten1965} Batten, A.~H., \& Hardie, R.~H.\ 1965, \aj, 70, 666 
\bibitem[Caga{\v s} \& Pejcha(2012)]{cagas2012} Caga{\v s}, P., \& Pejcha, O.\ 2012, \aap, 544, L3 
\bibitem[Carter et al.(2011)]{carter2011} Carter, J.~A., Fabrycky, D.~C., Ragozzine, D., et al.\ 2011, Science, 331, 562 
\bibitem[Catalano et al.(2002)]{catalano2002} Catalano, S., Biazzo, K., Frasca, A., \& Marilli, E.\ 2002, \aap, 394, 1009 
\bibitem[Dempsey et al.(1993)]{dempsey1993} Dempsey, R.~C., Linsky, J.~L., Fleming, T.~A., \& Schmitt, J.~H.~M.~M.\ 1993, \apjs, 86, 599
\bibitem[Derekas et al.(2011)]{derekas2011} Derekas, A., Kiss, L.~L., Borkovits, T., et al.\ 2011, Science, 332, 216 
\bibitem[Dotter et al.(2008)]{dotter2008} Dotter, A., Chaboyer, B., Jevremovi{\'c}, D., et al.\ 2008, \apjs, 178, 89
\bibitem[Fabricius et al.(2002)]{fabricius2002} Fabricius, C., H{\o}g, E., Makarov, V.~V., et al.\ 2002, \aap, 384, 180 
\bibitem[Flower(1996)]{flower1996} Flower, P. J. 1996, \apj, 469, 355
\bibitem[Gilliland \& Brown(1988)]{gilliland1988} Gilliland, R.~L., \& Brown, T.~M. 1988, \pasp, 100, 754
\bibitem[Graczyk et al.(2011)]{graczyk2011} Graczyk, D., Soszy{\'n}ski, I., Poleski, R., et al.\ 2011, \actaa, 61, 103 
\bibitem[Harmanec(1988)]{harmanec1988} Harmanec, P.\ 1988, Bulletin of the Astronomical Institutes of Czechoslovakia, 39, 329 
\bibitem[H{\o}g et al.(2000)]{hog2000} H{\o}g, E., Fabricius, C., Makarov, V.~V., et al.\ 2000, \aap, 355, L27 
\bibitem[Kim et al.(2007)]{kim2007} Kim, K.-M., Han, I., Valyavin, G.~G., et al.\ 2007, \pasp, 119, 1052 
\bibitem[Kim et al.(2001)]{kim2001} Kim, S.-L., Chun, M.-Y., Park, B.-G., et al.\ 2001, \aap, 371, 571 
\bibitem[Ko{\l}aczkowski et al.(2013)]{kolaczkowski2013} Ko{\l}aczkowski, Z., Mennickent, R.~E., Rivinius, T., \& Pietrzy{\'n}ski, G.\ 2013, \mnras, 429, 2852
\bibitem[Kurucz(1993)]{kurucz1993} Kurucz, R.\ 1993, ATLAS9 Stellar Atmosphere Programs and 2 km/s grid.~Kurucz CD-ROM No.~13.~ Cambridge, Mass.: Smithsonian Astrophysical Observatory 
\bibitem[Kwee \& van Woerden(1956)]{kwee1956} Kwee, K.~K., \& van Woerden, H.\ 1956, \bain, 12, 327
\bibitem[Lee et al.(2012a)]{bclee2012} Lee, B.-C., Han, I., Park, M.-G., Mkrtichian, D.~E., \& Kim, K.-M.\ 2012a, \aap, 546, A5 
\bibitem[Lee et al.(2008a)]{leecu2008} Lee, C.-U., Kim, S.-L., Lee, J.~W., et al.\ 2008a, \mnras, 389, 1630
\bibitem[Lee et al.(2007)]{lee2007} Lee, J. W., Kim, H.-I., \& Kim, S.-L. 2007, \pasp, 119, 1099
\bibitem[Lee et al.(2013a)]{jwlee2013} Lee, J.~W., Kim, S.-L., Lee, C.-U., et al.\ 2013a, \apj, 763, 74 
\bibitem[Lee et al.(2008b)]{lee2008} Lee, J.~W., Youn, J.-H., Kim, C.-H., Lee, C.-U., \& Kim, H.-I.\ 2008b, \aj, 135, 1523 
\bibitem[Lee et al.(2013b)]{lee2013} Lee, J. W., Youn, J.-H., Kim, S.-L., \& Lee, C.-U. 2013b, \aj, 145, 16
\bibitem[Lee et al.(2012b)]{lee2012} Lee, J.~W., Youn, J.-H., Kim, S.-L., Lee, C.-U., \& Hinse, T.~C.\ 2012b, \aj, 143, 95 
\bibitem[Lehmann et al.(2012)]{lehmann2012} Lehmann, H., Zechmeister, M., Dreizler, S., Schuh, S., \& Kanzler, R.\ 2012, \aap, 541, A105 
\bibitem[Lohr et al.(2013)]{lohr2013} Lohr, M.~E., Norton, A.~J., Kolb, U.~C., et al.\ 2013, \aap, 549, A86
\bibitem[Lucy(1967)]{lucy1967} Lucy, L.~B.\ 1967, \zap, 65, 89
\bibitem[Maceroni \& Rucinski(1997)]{maceroni1997} Maceroni, C., \& Rucinski, S.~M.\ 1997, \pasp, 109, 782 
\bibitem[Markwardt(2009)]{markwardt2009} Markwardt, C.~B.\ 2009, Astronomical Data Analysis Software and Systems XVIII, 411, 251
\bibitem[Mochnacki(1984)]{mochnacki1984} Mochnacki, S. W. 1984, \apjs, 55, 551
\bibitem[Ofir(2008)]{ofir2008} Ofir, A.\ 2008, Information Bulletin on Variable Stars, 5868, 1 
\bibitem[Pollacco et al.(2006)]{pollacco2006} Pollacco, D.~L., Skillen, I., Collier Cameron, A., et al.\ 2006, \pasp, 118, 1407
\bibitem[Ruci{\'n}ski(1969)]{rucinski1969} Ruci{\'n}ski, S.~M.\ 1969, \actaa, 19, 245
\bibitem[Sbordone(2005)]{sbordone2005} Sbordone, L.\ 2005, Memorie della Societa Astronomica Italiana Supplementi, 8, 61
\bibitem[Sbordone et al.(2004)]{sbordone2004} Sbordone, L., Bonifacio, P., Castelli, F., \& Kurucz, R.~L.\ 2004,  Memorie della Societa Astronomica Italiana Supplementi, 5, 93 
\bibitem[Schlafly \& Finkbeiner(2011)]{schlafly2011} Schlafly, E. F., \& Finkbeiner, D. P. 2011, \aj, 737, 103
\bibitem[Southworth et al.(2011)]{southworth2011} Southworth, J., et al. 2011, \mnras, 414, 2413
\bibitem[Stassun et al.(2012)]{stassun2012} Stassun, K.~G., Kratter, K.~M., Scholz, A., \& Dupuy, T.~J.\ 2012, \apj, 756, 47 
\bibitem[Tokovinin(1997)]{tokovinin1997} Tokovinin, A.~A.\ 1997, \aaps, 124, 75 
\bibitem[Torres(2010)]{torres2010} Torres, G. 2010, \aj, 140, 1158
\bibitem[Torres(2013)]{torres2013} Torres, G.\ 2013, Astronomische Nachrichten, 334, 4 
\bibitem[Van Hamme(1993)]{van1993} Van Hamme, W. \ 1993, \aj, 106, 2096
\bibitem[Voges et al.(1999)]{voges1999} Voges, W., Aschenbach, B., Boller, T., et al.\ 1999, \aap, 349, 389 
\bibitem[Wilson \& Devinney(1971)]{wilson1971} Wilson, R. E., \& Devinney, E. J. 1971, \apj, 166, 605
\bibitem[Zacharias et al.(2013)]{zacharias2013} Zacharias, N., Finch, C.~T., Girard, T.~M., et al.\ 2013, \aj, 145, 44 
\end{thebibliography}
\end{document}